\documentclass[proceedings]{JHEP37}
\usepackage{eurosym}
\usepackage{amssymb}
\usepackage{amsfonts,cite}
\usepackage{amsmath}
\usepackage{graphicx}
\usepackage{multirow}
\usepackage{epsfig}
\usepackage[title]{appendix}
\DeclareMathOperator{\arctanh}{arctanh}

\newbox\mybox

\newcommand\fverb{\setbox\mybox=\hbox\bgroup\verb}
\newcommand\fverbdo{\egroup\medskip\noindent\fbox{\unhbox\mybox}\ }
\newcommand\fverbit{\egroup\item[\fbox{\unhbox\mybox}]}
\conference{LR-invariants from 2D point transformations}
\abstract{We construct Lewis-Riesenfeld invariants from two dimensional point transformations for two oscillators that are coupled to each other in space in a PT-symmetrical and  time-dependent fashion. The non-Hermitian Hamiltonian of the model is conveniently expressed in terms of generators of the symplectic $sp(4)$ Lie algebra. This allows for an alternative systematic approach to find Lewis-Riesenfeld invariants leading to a set of coupled differential equations that we solve by using time-ordered exponentials. We also demonstrate that point transformations may be utilized to directly construct time-dependent Dyson maps from their respective time-independent conterparts in the reference system.}

\title{Lewis-Riesenfeld invariants for PT-symmetrically coupled oscillators from two dimensional point transformations and Lie algebraic expansions}
\author{Andreas Fring and Rebecca Tenney\\
 Department of Mathematics, City, University of London, Northampton Square,\\ London EC1V 0HB, UK \\
a.fring@city.ac.uk, rebecca.tenney@city.ac.uk}

\input{tcilatex}
\begin{document}

\section{Introduction}
While in general time-independent quantum non-Hermitian Hamiltonian systems are fairly well understood \cite{Urubu,Bender:1998ke,Alirev,PTbook}, and have also found many applications, their time-dependent versions pose a more technical challenge. These type of systems are especially interesting as their exceptional points do not signify a transition from a physical to a non-physical region as is typical in a time-independent setting, but merely mark the boundary in parameter space between different types of qualitative behaviour \cite{AndTom3,fring2021perturb,fring2022time,huang2022solvable}. In turn this feature gives rise to new types of physical effects observed for instance in their associated von-Neumann entropies \cite{fring2019eternal,frith2020exotic} and instantaneous energy spectra \cite{fring2021perturb}. Naturally to solve the governing time-dependent Schr\"odinger equation (TDSE) is more challenging and due to their non-Hermitian nature one needs to construct in addition a time-dependent metric operator to obtain a meaningful quantum theory \cite{CA,time1,time6,BilaAd,time7,fringmo,fring2022intro}. Both tasks can be facilitated by the use of Lewis-Riesenfeld (LR) invariants \cite{Lewis69}.     

In general LR-invariants have proven to be extremely useful tools in the endeavour to obtain exact, or very good approximate solutions \cite{fring2020time}, to the time-dependent Schr\"odinger equation (TDSE), as they reduce the latter to a simpler eigenvalue equation with time-independent eigenvalues where time simply plays the role of a standard parameter. While this feature holds for time-dependent Hermitian \cite{pedrosa1997exact} as well as for  time-dependent non-Hermitian systems \cite{maamache2017pseudo,khantoul2017invariant,AndTom4,cen2019time,fring2021exactly}, in the latter context LR-invariants also facilitate the construction of the Dyson map $\eta(t)$, which in turn factorises the metric operator $\rho(t)=\eta^\dagger(t) \eta(t)$ that is essential in setting up a well-defined quantum theory for the non-Hermitian system. Using the LR-invariant $I_H(t)$ allows to bypass solving the time-dependent Dyson equation, which usually boils down to a set of coupled differential equations. Instead the task is reduced to mapping the non-Hermitian LR-invariant associated with a non-Hermitian Hamiltonian $H(t)$ to a Hermitian one related to a Hermitian Hamiltonian $h(t)$ by means of a similarity transformation $I_h(t)= \eta I_H(t) \eta^{-1}$. Moreover, it was recently shown \cite{fring2022time} that when constraining the invariant $I_H(t)$ to be an involution and in addition require it to commute with the $\cal{PT}$-operator, i.e. a simultaneous parity and time-reversal, it becomes identical to the time-dependent ${\cal{C}}(t)$-operator, which is directly related to the time-dependent metric as $\rho(t)={\cal{PC}}(t)$.  

Thus, for the various reasons mentioned above it is important to have effective means to construct LR-invariants in an efficient manner. The standard construction relies on an inspired Ansatz for the invariant leading usually to a set of complicated coupled differential equations. Following \cite{fring2022time} we demonstrate here how these equations can be solved in an efficient manner using time-ordered exponentials. In the main part of the paper we follow an alternative route and construct invariants from point transformations. Previously this was carried out for one-dimensional Hermitian \cite{zelaya2020quantum} and non-Hermitian systems \cite{BeckyAndPoint}. Here we generalise this approach to the more complicated two-dimensional setting. Moreover we show that one can use the point transformations to directly construct the time-dependent Dyson map for a non-Hermitian time-dependent Hamiltonian as indicated in \cite{beckythesis}.

As a concrete model we discuss two harmonic oscillators that are coupled to each other in space in a $\cal{PT}$-symmetric fashion with explicitly time-dependent coefficient functions. The Hamiltonian can be expressed conveniently in terms of generators of the symplectic $sp(4)$ Lie algebra. 

Our manuscript is organised as follows: In Section 2 we recall some well known features of the symplectic $sp(4)$ Lie algebra, in particular two representations, that are essential to set up our model. In Section 3 we construct the LR-invariants for our non-Hermitian model. In the first part we express the invariants in terms of the $sp(4)$ Lie algebraic generators and solve the set of coupled differentials equations that arise from substituting this Ansatz into the LR equation by means of time-ordered exponentials. In the second part we construct the invariants from two dimensional point transformations using suitable reference Hamiltonians. Moreover, by acting with the point transformation on the more easily obtainable Dyson map of the time-independent reference Hamiltonian, we construct their corresponding time-dependent Dyson map and time-dependent Hermitian Hamiltonian counterpart. Our conclusions are stated in Section 4. 

\newpage

\section{Mathematical preliminaries - the symplectic $sp(4)$-algebra}
It is almost always convenient to express the model to be studied in terms a well-defined Lie algebra. In our case we take this to be the symplectic Lie algebra $sp(4,\mathbb{F})$ over the field $\mathbb{F}=\mathbb{R}$ or $\mathbb{F}=\mathbb{C}$. Its generators are defined as the set of $(4\times 4)$-matrices $M$ satisfying
\begin{equation}
	\Omega M  + M^\intercal \Omega =0, \qquad \text{with} \,\, \Omega = \left(
	\begin{array}{cc}
		0 & \mathbb{I} \\
		-\mathbb{I} & 0 \\
	\end{array}
	\right) , \label{sympldef}
\end{equation}
where $\mathbb{I}$ denotes the $(2 \times 2)$-unit matrix, see for instance chapter 16 in \cite{fulton2013rep}. One may find the following Hermitian matrices solutions to (\ref{sympldef}) that commute with $\Omega$
\begin{eqnarray}
	J_0 &=& \frac{\imath}{2} \left(
	\begin{array}{cc}
		0 & \mathbb{I} \\
		-\mathbb{I} & 0 \\
	\end{array}
	\right), \quad J_1 = \frac{\imath}{2} \left(
	\begin{array}{cc}
		0 & \sigma_1 \\
		- \sigma_1 & 0 \\
	\end{array}
	\right), \quad J_2 = \frac{\imath}{2} \left(
	\begin{array}{cc}
		\sigma_2 & 0 \\
		0 & \sigma_2 \\
	\end{array}
	\right), \quad J_3 = \frac{\imath}{2} \left(
	\begin{array}{cc}
		0 & \sigma_3 \\
		- \sigma_3 & 0 \\
	\end{array}
	\right). \qquad \label{Matrep1}
\end{eqnarray}
Here the $\sigma_i$ with $i=1,2,3$ are standard $(2 \times 2)$-Pauli matrices. The anti-Hermitian matrices
\begin{eqnarray}
	Q_1 &=& \frac{\imath}{2} \left(
	\begin{array}{cc}
		-\sigma_3 & 0 \\
		0 & \sigma_3 \\
	\end{array}
	\right), \quad Q_2 = \frac{\imath}{2} \left(
	\begin{array}{cc}
		0 & \mathbb{I} \\
		\mathbb{I} & 0 \\
	\end{array}
	\right), \quad Q_3 = \frac{\imath}{2} \left(
	\begin{array}{cc}
		\sigma_1 & 0 \\
		0 & -\sigma_1 \\
	\end{array}
	\right), \\
	K_1 &=&  \frac{\imath}{2} \left(
	\begin{array}{cc}
		0 & \sigma_3 \\
		\sigma_3 & 0 \\
	\end{array}
	\right),  \quad K_2 = \frac{\imath}{2} \left(
	\begin{array}{cc}
		\mathbb{I} & 0 \\
		0 & -\mathbb{I} \\
	\end{array}
	\right), \quad K_3 =  -\frac{\imath}{2} \left(
	\begin{array}{cc}
		0 & \sigma_1 \\
		\sigma_1 & 0 \\
	\end{array}
	\right), \label{Matrep2}
\end{eqnarray}
anti-commute with $\Omega$ and also solve (\ref{sympldef}), see e.g. \cite{han1993symmetries}. We notice that these matrices are related to the standard gamma matrices occurring in the Dirac equation as $J_1=\imath/2 \gamma_1$, $J_2=\imath/2 \gamma_2$, $J_3=\imath/2 \gamma_3$, $K_2=\imath/2 \gamma_0$, $Q_2=\imath/2 \gamma_5$, reflecting the fact that the $Sp(4)$ group is isomorphic to the $O(3,2)$ de Sitter group \cite{dirac1963remarkable}. 

The ten generator  $J_0,J_i,Q_i,K_i$, for $i=1,2,3$ obey the commutation relations 
\begin{align}
	\left[ J_i ,J_j \right]&= \imath \varepsilon_{ijk} J_k, \quad & \left[ J_i ,K_j \right]& = \imath \varepsilon_{ijk} K_k, \quad & \left[ J_i ,Q_j \right]&= \imath \varepsilon_{ijk} Q_k, \label{co1}\\
	\left[ J_i ,J_0 \right]&= 0, \quad & \left[ K_i ,J_0 \right]& = \imath Q_i, \quad & \left[ Q_i ,J_0 \right]&= - \imath K_i, \\
		\left[ K_i ,K_j \right]&= -\imath \varepsilon_{ijk} J_k, \quad & \left[ Q_i ,Q_j \right]& = -\imath \varepsilon_{ijk} J_k, \quad & \left[ K_i ,Q_j \right]&= \imath \delta_{ij} J_0, \label{co3}
\end{align}
which is easily verified for the matrix representation (\ref{Matrep1})-(\ref{Matrep2}), with $\varepsilon_{ijk}$ denoting the standard Levi-Civita tensor and $\delta_{ij}$ the Kronecker delta symbol. We observe that this algebra remains invariant under the following antilinear symmetries
\begin{eqnarray}
	{\cal{PT}}: &&\!\!  J_0 \rightarrow J_0, \,\, J_1 \rightarrow -J_1, \,\, J_2 \rightarrow J_2, \,\, J_3 \rightarrow J_3, \,\, Q_1 \rightarrow -Q_1, 
	\,\, Q_2 \rightarrow Q_2, 	\,\, Q_3 \rightarrow Q_3, \quad \label{PTalg} \\
	&& \!\! K_1 \rightarrow K_1, 
	\,\, K_2 \rightarrow -K_2, 	\,\, K_3 \rightarrow -K_3, \,\, \imath \rightarrow - \imath , \notag \\
		\tilde{{\cal{PT}}}: && \!\! J_0 \rightarrow -J_0, \,\, J_1 \rightarrow J_1, \,\, J_2 \rightarrow J_2, \,\, J_3 \rightarrow -J_3, \,\, Q_1 \rightarrow -Q_1, 
	\,\, Q_2 \rightarrow -Q_2, 	\,\, Q_3 \rightarrow Q_3, \qquad \label{PTalg2} \\
	&& \!\!  K_1 \rightarrow -K_1, 
	\,\, K_2 \rightarrow -K_2, 	\,\, K_3 \rightarrow K_3, \,\, \imath \rightarrow - \imath , \notag 	
\end{eqnarray} 
Alternatively we may express the generators in terms of the coordinates $x,y$ with corresponding momenta $p_x,p_y$ and non-vanishing commutators $[x,p_x]=[y,p_y]=\imath$ as
\begin{eqnarray}
	J_0 &=& \frac{1}{4} \left( p_x^2 +p_y^2 +  x^2 +  y^2   \right), \quad \label{rep21}\\
	J_1 &=& \frac{1}{2} \left( x y+ p_x p_y    \right), \quad
	J_2 = \frac{1}{2} \left(x p_y - y p_x   \right), \quad 
	J_3 = \frac{1}{4} \left( p_x^2 - p_y^2 + x^2  - y^2   \right), \quad\\
	Q_1 &=& \frac{1}{2} \left( y p_y -  x p_x   \right), \quad 
	Q_2 = \frac{1}{4} \left( p_x^2  + p_y^2 - x^2 - y^2   \right), \quad
	Q_3 = \frac{1}{2} \left( x p_y +  y p_x   \right), \quad \\
	K_1 &=& \frac{1}{4} \left( p_x^2  - p_y^2 - x^2 - y^2   \right), \quad
	K_2 = \frac{1}{2} \left( x p_x +   p_y y  \right), \quad
	K_3 = \frac{1}{2} \left( x y -  p_x p_y   \right). \label{rep22}
\end{eqnarray}
 One verifies that the generators expressed in this manner also satisfy the relations (\ref{co1})-(\ref{co3}). This Hermitian representation was already known to Dirac, see \cite{dirac1963remarkable}. In this case the ${\cal PT}$-symmetry (\ref{PTalg}) corresponds to the mappings
\begin{equation}
	{\cal{PT}}_\pm: x \rightarrow \pm x, \quad y \rightarrow \mp y \quad 
	 p_x \rightarrow \mp p_x, \quad  p_y \rightarrow \pm p_y, \quad \imath \rightarrow - \imath .
\end{equation} 
Identifying the parity operator as ${\cal P}= e^{\imath \pi J_3}$ and the time-reversal operator ${\cal T}$ as complex conjugation, the adjoint action of ${\cal PT}$ on all the generators leads precisely to the mapping in (\ref{PTalg}).

\section{Time-dependent PT-symmetrically coupled oscillators from $sp(4)$}
Let us now introduce the non-Hermitain model we will be investigating, the time-dependent version of two oscillators coupled to each other in a complex ${\cal PT}$-symmetrical fashion in space
\begin{equation}
	H = \frac{a(t)}{2} \left( p_x^2 + p_y^2 \right) + \frac{1}{2} \left[  \omega_x(t) x^2 + \omega_y(t) y^2 \right] + i \lambda(t)x y. \label{PTcoupledHO}
\end{equation}
For the time-independent scenario, i.e. $ \dot{a}= \dot{\omega}_x=\dot{\omega}_y=\dot{\lambda}= 0$, this system was previously studied in \cite{MandalMY,beygi2015coupled,AndTom4}. Given the representation of the $sp(4)$ Lie algebra (\ref{rep21})-(\ref{rep22}), we can express this Hamiltonian entirely in terms its generators  
\begin{equation}
	H = \frac{a(t)}{2} \left( J_0 +Q_2 \right) + \frac{\Omega_+(t)}{2} \left(  J_0 - Q_2 \right) + \frac{\Omega_-(t)}{2} \left(  J_3 - K_1 \right) 
	 + i \lambda(t) \left(  J_1 +K_3  \right) , 
\end{equation}
with $\Omega_\pm(t) := \omega_x(t) \pm \omega_y(t) $. For the $(4 \times 4)$-matrix version of the Hamiltonian $H(t)$ the four instantaneous eigenvalues are easily computed to
\begin{equation}
     \epsilon_{\pm}^{\pm } = \pm \frac{1}{2} \left[ a(t) \Omega_+^2(t) \pm a(t) \sqrt{ \Omega_+^2(t) -4 \lambda^2(t)  }       \right]^{1/2}   . 
\end{equation}
It should be stressed at this point that these are not the energy eigenvalues of the time-dependent system, since $H$ is non-Hermitian, see e.g. \cite{fringmo}. For the time-independent case this agrees with the values found for the ground state in \cite{MandalMY,beygi2015coupled,AndTom4}, so that in analogy to that scenario we will speak of a ${\cal PT}$-symmetric and spontaneously broken regime.

We identify ${\cal P} = 2 J_3$ as the parity operator from the relation ${\cal P} H {\cal P} = H^\dagger$ and the time-reversal operator ${\cal{T}}$ as standard complex conjugation. 
\subsection{LR-invariants from a Lie algebraic expansion}
The defining relation for the Lewis-Riesenfeld invariants $I_H(t)$ is that it satisfies the Lewis-Riesenfeld equation introduced in \cite{Lewis69} 
\begin{equation}
	\imath \hbar \partial_t I_H(t)  - [H,I_H(t)]=0 . \label{LRdef}
\end{equation}	
Most attempts to explicitly construct $I_H(t)$ by solving (\ref{LRdef}) start by making a suitable general Ansatz for the invariant. Considering a system with an underlying Lie algebra ${\bf g}$ it is suggestive to simply expand the invariant in terms of the Lie algebraic generators $ X_i$ for $i=1,\ldots ,\text{rank}$  ${\bf g} $, as $I_H = \sum_{i=1}^{10} c_i(t) X_i$. Here it will be slightly more convenient to assume
\begin{eqnarray}
	I_H &=& c_1(t) \left(  Q_1 -K_2  \right) + 
	c_2(t) \left(  Q_1 +K_2  \right) +
	c_3(t) \left( J_2 + Q_3  \right) +
	c_4(t) \left(  J_2 - Q_3  \right) \label{AnsatzI} \\
	&& \!\!\! +
	c_5(t) \left(  J_1 +K_3  \right) +
	c_6(t) \left(  J_1 - K_3 \right) +
	c_7(t) \left( J_0 - J_3 + K_1 + Q_2  \right) \notag \\
	&& \!\!\! +c_8(t) \left(  J_0 + J_3 - K_1 - Q_2  \right) +
	c_9(t) \left(  J_0 + J_3 + K_1 + Q_2  \right) +
	c_{10}(t) \left(  J_0 - J_3 - K_1 + Q_2  \right). \notag
\end{eqnarray}
The  Ansatz (\ref{AnsatzI}) is a priori not obvious but chosen in hindsight to generate simpler constraints thereafter when substituting it into the Lewis-Riesenfeld equation (\ref{LRdef}). In doing so we obtain the ten constraints for the time-dependent coefficient functions
\begin{eqnarray}
\dot{c}_1 &=& a c_8- 2 \omega_x c_9-\imath \lambda c_6 , \quad 
\dot{c}_2 = 2 \omega_y c_{10}- a c_7+\imath \lambda c_6 , \quad
\dot{c}_3 = \omega_x c_6 - \frac{1}{2} a c_5 + 2 \imath \lambda c_{10} , \notag \\
\dot{c}_4 &=& \frac{1}{2} a c_5 - \omega_y c_6 - 2 \imath \lambda c_{9} , \quad
\dot{c}_5 =  \omega_y c_3 -  \omega_x c_4 +  \imath \lambda \left( c_2 -c_1 \right), \quad
\dot{c}_6 = \frac{1}{2} a  \left( c_4 -  c_3 \right), \label{constraint} \quad\\
\dot{c}_7 &=&  \omega_y c_2 -\imath \lambda c_4, \quad
\dot{c}_8 = - \omega_y c_1 +\imath \lambda c_3, \quad
\dot{c}_9 = \frac{1}{2} a c_1, \quad \dot{c}_{10} = - \frac{1}{2} a c_2. \notag
\end{eqnarray} 
Thus, given the functions $a(t), \omega_x(t), \omega_y(t), \lambda(t)$, we have to solve the system (\ref{constraint}) for the unknown time-dependent coefficient functions $c_i(t)$, with $i=1,\ldots,10$. 
\subsubsection{Solutions from time-ordered exponentials}
Next we solve the system of coupled first order differential equations (\ref{constraint}) by means of time-ordered exponentials following \cite{fring2022time}. For this purpose we first re-write (\ref{constraint}) as a matrix equation
\begin{equation}
	\partial_t \vec{c} = M \vec{c}, \qquad   \text{with} \,\, M= \left(
	\begin{array}{cccccccccc}
		0 & 0 & 0 & 0 & 0 & -i \lambda  & 0 &  \, a \,  & -2 \omega_x & 0 \\
		0 & 0 & 0 & 0 & 0 & i \lambda  & \, a \,  & 0 & 0 & 2 \omega_y \\
		0 & 0 & 0 & 0 & -\frac{a}{2} & \omega_x & 0 & 0 & 0 & 2 i \lambda  \\
		0 & 0 & 0 & 0 & \frac{a}{2} & -\omega_y & 0 & 0 & -2 i \lambda  & 0 \\
		-i \lambda  & i \lambda  & \omega_y & -\omega_x & 0 & 0 & 0 & 0 & 0 & 0 \\
		0 & 0 &-\frac{a}{2} & \frac{a}{2} & 0 & 0 & 0 & 0 & 0 & 0 \\
		0 & \omega_y & 0 & -i \lambda  & 0 & 0 & 0 & 0 & 0 & 0 \\
		-\omega_x & 0 & i \lambda  & 0 & 0 & 0 & 0 & 0 & 0 & 0 \\
		\frac{a}{2} & 0 & 0 & 0 & 0 & 0 & 0 & 0 & 0 & 0 \\
		0 & -\frac{a}{2} & 0 & 0 & 0 & 0 & 0 & 0 & 0 & 0 \\
	\end{array}
	\right) ,
\end{equation}
so that the general solution can be expressed in terms of time-ordered exponential as
\begin{eqnarray}
	\vec{c}(t)  &=& T \exp\left[  \int_0^t M(s) ds  \right] \vec{c}(0)  \label{gensolM} \\
	&=&  \sum_{n=0}^{\infty}  T  \left[ \int_0^t M(t_1) dt_1 \int_0^{t_1} M(t_2) dt_2 \ldots \int_0^{t_{n-1}} M(t_{n-1}) dt_n \right] \vec{c}(0) , \label{timeord}
\end{eqnarray}
with $t> t_1> \ldots t_n >0$ and $T$ denoting the time-ordering operator. When the matrix $M(t)$ commutes for different times, i.e. $[M(t),M(t')]=0$ for $t\neq t'$, this expression can be evaluated directly, as the time-ordering can be dropped so that (\ref{gensolM}) just becomes a matrix exponential
\begin{equation}
	\vec{c}(t)  = \exp\left[  \int_0^t M(s) ds  \right] \vec{c}(0)  =
	\sum_{n=0}^{\infty}  \frac{1}{n!} \left[  \int_0^t M(s) ds  \right]^n \vec{c}(0) . \label{matex}
\end{equation}
The commutativity requirement on $M(t)$ imposes the functions $a(t), \omega_x(t), \omega_y(t), \lambda(t)$ to be proportional to each other. Indeed, when making this assumption we find some general solutions to the Lewis-Riesenfeld equation (\ref{LRdef}) in this manner. However, the solutions are rather involved including also the 10 constants from the initial condition $\vec{c}(0)$ and we will therefore not present here the generic expression.

Instead we make a few concrete and well-motivated choices that lead to simpler expressions. First of all we take the proportionality to be $a(t)=\lambda(t), \omega_x(t)= \alpha \lambda(t), \omega_y(t)=\lambda(t)$, leaving us with an arbitrary constant $\alpha$. Moreover, keeping in mind that we would like to employ the invariants in the construction of metric operators, we make use of the observation made in \cite{fring2022time}, that restricted Lewis-Riesenfeld invariants become time-dependent ${\cal C}(t)$-operators, which when multiplied with the parity ${\cal P}$ operator becomes the metric operator $\rho(t) = {\cal P C}(t)$. The key restriction to be imposed is the involution property $I_H^2 = \mathbb{I}$. Let us now see how this is implemented for the invariant (\ref{AnsatzI}), which when expressing the $sp(4)$-generators in the matrix representation (\ref{Matrep1})-(\ref{Matrep2}) acquires the form
\begin{equation}
I_H(t) = \imath \left(
\begin{array}{rrrr}
	-c_1(t) & -c_4(t) & 2 c_9(t) & c_6(t) \\
	c_3(t) & c_2(t) & c_6(t) & 2 c_{10}(t) \\
	-2 c_8(t) & -c_5(t) & c_1(t) & -c_3(t) \\
	-c_5(t) & -2 c_7(t) & c_4(t) & -c_2(t) \\
\end{array}
\right) .
\end{equation}
This expression squares to $\mathbb{I}$ when implementing the following constraints
\begin{eqnarray}
	c_1 &=& \sqrt{4 c_8 c_9 + \chi_+ -1}, \qquad \quad \,\,\,	c_2 = \frac{2 \left(c_4 c_6 c_8-c_3 c_5 c_9\right)}{\chi _+} + \frac{ \chi _-}{\chi _+} c_1, \label{const1}\\
		c_7 &=& \frac{c_1 c_4 c_5}{\chi _+}-\frac{c_8 c_4^2+c_9 c_5^2 }{\chi _+}, \qquad c_{10} = -\frac{c_1 c_3 c_6}{\chi _+}-\frac{c_9 c_3^2+ c_8 c_6^2 }{\chi _+}, \label{const2}
\end{eqnarray}
where we abbreviated $\chi_{\pm} = c_3 c_4 \pm c_5 c_6 $. With these constraints we also obtain $\det I_H = 1$. Next we chose an initial condition that respects these constraints also at the time $t=0$ as $\vec{c}(0)=(0,0,1,1,0,0,0,0,0,0)$. Evaluating the matrix exponential (\ref{matex}) by acting on this vector as the chosen initial condition yields the solution
\begin{eqnarray}
c_1(t) &=& c_2(t) = \frac{\imath}{2 \sqrt{1 + \alpha}} \left( C_- - C_+   \right), \\ 
c_3(t)&=& \frac{\sqrt{1+\alpha } + \alpha
	}{2 \sqrt{1+\alpha }} C_-  +\frac{\sqrt{1+\alpha }-\alpha  }{2 \sqrt{1+\alpha }} C_+, \,\, 
c_4(t) =   \frac{\sqrt{1+\alpha } +1
}{2 \sqrt{1+\alpha }} C_-  +\frac{\sqrt{1+\alpha }-1  }{2 \sqrt{1+\alpha }} C_+ , \,\,\,\, \qquad \\
c_5(t)&=& \frac{1-\alpha  }{a_+ }S_+  + \frac{1-\alpha  }{a_-} S_-, \,\,
c_6(t) = \frac{\alpha -1  }{2 \sqrt{1+\alpha}a_+ } S_+ + \frac{1-\alpha  }{ 2 \sqrt{1+\alpha} a_-} S_-,
      \\
c_8(t) &=& -c_7(t) = \frac{\imath S_+}{ a_+}+\frac{ \imath S_-}{ a_-}, \quad
c_9(t) = -c_{10}(t) = \frac{\imath S_-}{2  \sqrt{1+\alpha } a_-}-\frac{\imath S_+}{2  \sqrt{1+\alpha } a_+ }, \qquad\,\,\,\,
\end{eqnarray}
where
\begin{equation}
	C_{\pm} = \cos \left[ \frac{a_\pm}{2} \int^t \! \lambda (s) \, ds\right], \,\,	S_{\pm} =  \sin \left[\frac{a_\pm}{2} \int^t \! \lambda (s) \, ds\right], \,\, a_\pm = \sqrt{2} \sqrt{1+\alpha \pm 2 \sqrt{\alpha +1}} .
\end{equation}
We convinced ourselves that these solutions for the time-dependent coefficient function do indeed satisfy the constraints (\ref{const1}) and (\ref{const2}).

\subsection{LR-invariants from two dimensional point transformations}

\subsubsection{The general scheme}

We will now utilise two dimensional point transformations to construct LR-invariants and time-dependent Dyson maps for the above system. In general, point transformation are canonical transformations in time and configuration coordinate space extended to a phase-space transformation. Here we define the point transformations $\Gamma$ as the map that maps the TDSE for a two-dimensional time-independent reference Hamiltonian $H_{0}(\chi,v )$, depending on the coordinates $\chi,v$, to the TDSE for a two-dimensional time-dependent non-Hermitian target Hamiltonian $H(x,y,t)$
\begin{eqnarray}
	\Gamma :  H_{0}(\chi,v )\Psi (\chi ,v,\tau )=\imath \hbar \partial _{\tau }\Psi
	(\chi ,v,\tau ) && \rightarrow H(x,y,t)\Phi
	(x,t)=\imath \hbar \partial _{t}\Phi (x,y,t),  \label{gamma1} \\ 
	 \left[\chi ,v,\tau ,\Psi (\chi,v,\tau )\right] && \mapsto \left[ x,y,t,\Phi (x,y,t)\right] .  \label{gamma}
\end{eqnarray}%
The wave functions $\Psi $ and $\Phi $ are implicit functions of the coordinates and time $\chi,v $,$%
\tau $ and $x,y$,$t$, respectively, defined by the associated TDSEs. The variables $\chi,v $, $\tau $, $\Psi $ are understood in
general as functions  $P$, $Q$, $R$, $S$ of the new set of independent variables $x$,  $y$, $t$, $%
\Phi $
\begin{equation}
	\chi =P(x,y,t,\Phi ),\text{\qquad }v =Q(x,y,t,\Phi ),\text{\qquad }\tau =R(x,y,t,\Phi ),\qquad \Psi =S(x,y,t,\Phi
	).  \label{genpoint}
\end{equation}%
For concrete systems one may be forced naturally to drop some of the dependences or simply relax them for convenience. 

The key feature is now that the point transformation $\Gamma$ defined via (\ref{gamma1}) can be applied to various other quantities that otherwise need to be constructed in a more complicated and involved manner. In particular, when acting solely on the Hamiltonian $H_{0}$, the map leads to the LR-invariant $I_H(x,y,t)$ for the target system we are interested in. Moreover, when acting on the time-independent Dyson map for the reference Hamiltonian we obtain directly the time-dependent Dyson map for the time-dependent target Hamiltonian.    

We summarise the relations between various types of Schr\"odinger equations and Lewis-Riesenfeld eigenvalue equations as follows:
\[
\begin{array}{ccc}
	\fbox{$H_{0}(\chi,v )\Psi (\chi ,v,\tau )=\imath \hbar \partial _{\tau }\Psi
		(\chi ,v,\tau )$} & \overset{\Gamma }{\longrightarrow } & \fbox{$H(x,y,t)\Phi
		(x,t)=\imath \hbar \partial _{t}\Phi (x,y,t)$} \\ 
	\begin{array}{c}
		\updownarrow  \\ 
		\Psi (\chi ,v,\tau )=e^{-\imath \tau E/\hbar }\tilde{\Psi}(\chi,v ) \\ 
		\updownarrow  \\ 
	\end{array}
	&  & 
	\begin{array}{c}
		\updownarrow  \\ 
		\Phi (x,y,t)=e^{\imath \alpha (t)/\hbar }\tilde{\Phi}(x,y,t) \\ 
		\updownarrow  \\ 
	\end{array}
	\\ 
	\fbox{$H_{0}(\chi,v )\tilde{\Psi}(\chi,v)=E\tilde{\Psi}(\chi,v )$} & 
	~~~\overset{\Gamma }{\longrightarrow }~~~ & \fbox{$I_{H}(x,y,t)\tilde{\Phi}%
		(x,t)=\Lambda \tilde{\Phi}(x,y,t)$} \\ 
	\begin{array}{c}
		\updownarrow  \\ 
		\begin{array}{c}
			h_{0}(\chi,v )=\eta H_{0}(\chi,v )\eta ^{-1} \\ 
			\tilde{\Psi}(\chi,v  )=\eta \tilde{\psi}(\chi,v )%
		\end{array}
		\\ 
		\updownarrow  \\ 
	\end{array}
	&  & 
	\begin{array}{c}
		\updownarrow  \\ 
		\begin{array}{c}
			I_{h}(x,y,t)=\eta (t)I_{H}(x,t)\eta ^{-1}(t) \\ 
			\tilde{\Phi}(x,y,t)=\eta(t) \tilde{\phi}(x,y,t)%
		\end{array}
		\\ 
		\updownarrow  \\ 
	\end{array}
	\\ 
	\fbox{$h_{0}(\chi,v )\tilde{\psi}(\chi  )=E\tilde{\psi}(\chi,v  )$} & 
	\overset{\Gamma }{\longrightarrow } & \fbox{$I_{h}(x,t)\tilde{\phi}%
		(x,y,t)=\Lambda \tilde{\phi}(x,y,t)$} \\ 
	\begin{array}{c}
		\updownarrow  \\ 
		\psi (\chi ,v,\tau )=e^{-\imath \tau E/\hbar }\tilde{\psi}(\chi,v ) \\ 
		\updownarrow  \\ 
	\end{array}
	&  & 
	\begin{array}{c}
		\updownarrow  \\ 
		\phi (x,y,t )=e^{\imath\alpha(t)/\hbar }\tilde{\phi}(x,y) \\ 
		\updownarrow  \\ 
	\end{array}
	\\ 
	\fbox{$h_{0}(\chi,v )\psi (\chi ,v,\tau )=\imath \hbar \partial _{\tau }\psi
		(\chi ,v,\tau )$} & \overset{\Gamma }{\longrightarrow } & \fbox{$h(x,y,t)\phi
		(x,y,t)=\imath \hbar \partial _{t}\phi (x,y,t)$}%
\end{array}%
\]
 Let us briefly explain the diagram by descending down the first column starting in the top left corner with TDSE for the time-independent non-Hermitian reference Hamiltonian $H_0(\chi,v)$ that is mapped by factorization of the wave function into the time-independent Schr\"odinger equations (TISE). In the next step we carry out a similarity transformation using the time-independent Dyson map to obtain the time-independent Schr\"odinger equation for the Hermitian Hamiltonian $h_0(\chi,v)$, which again by factorization of the wave function is subsequently mapped to the TDSE for the time-independent Hermitian Hamiltonian $h_0(\chi,v)$.

In the second column we map the TDSE for the time-dependent non-Hermitian target Hamiltonian $H(x,y,t)$ to the eigenvalue equation for the non-Hermitian LR-invariant $I_H(x,y,t)$ with time-independent eigenvalue $\Lambda$ by factorising off the LR-phase from the wave function. The eigenvalue equation for the Hermitian LR-invariant $I_h(x,y,t)$ is then obtained by means of a similarity transformation using a time-dependent Dyson map, which when extracting again the LR-phase is converted into the TDSE for the time-dependent Hermitian Hamiltonian $h(x,y,t)$.   

All steps in each of the two columns have been successfully carried out for a number of explicit systems. However, in comparison the calculations in the second column are considerably more complicated than those in the first. Thus, the idea is to bypass these steps by means of a point transformation $\Gamma$ relating in the diagram the various equations  horizontally. We are especially interested in the construction of the non-Hermitian LR-invariant $I_H(x,y,t)$ from the time-independent Hamiltonian $H_0(\chi,v)$, the time-dependent Dyson map $\eta(t)$ from the time-independent Dyson map $\eta$ and the Hermitian Hamiltonian $h(x,y,t)$ from the action of $\Gamma$ on the TDSE satisfied by $h_{0}(\chi,v )$ 
\begin{eqnarray}
	\Gamma: && H_0(\chi) \rightarrow I_H(x,t), \\
	\Gamma: && \eta \rightarrow \eta(t), \\ 
	\Gamma: && h_{0}(\chi,v )\psi (\chi ,v,\tau )=\imath \hbar \partial _{\tau }\psi
	(\chi ,v,\tau ) \quad \rightarrow \quad  h(x,y,t)\phi
	(x,y,t)=\imath \hbar \partial _{t}\phi (x,y,t), \quad
\end{eqnarray}
The second mapping follows by decomposing $\Gamma: \tilde{\Psi}(\chi,v  ) \rightarrow \tilde{\Phi}(x,y,t )$ into $\Gamma: \tilde{\psi}(\chi,v  ) \rightarrow \tilde{\phi}(x,y,t )$ and the remainder $\Gamma: \eta \rightarrow \eta(t)$.  
Let us now turn to our concrete model.
\subsubsection{Point transformation $\Gamma$ for a general target Hamiltonian}
We consider here a slightly modified version of the Hamiltonian in (\ref{PTcoupledHO}) given by
\begin{equation}\label{target}
	H(x,y,t) = \frac{a(t)}{2}\left(p_x^2+x^2\right)+\frac{b(t)}{2}\left(p_y^2+y^2\right)+\imath \lambda(t)xy,
\end{equation}
which in terms of the generators of the $sp(4)$-algebra is expressed as
\begin{equation}
	H = a(t)\left(J_3+J_0\right)+b(t)\left(J_0-J_3\right)+\imath \lambda(t)\left(J_1+K_3\right).
\end{equation}
We will now employ two dimensional point transformations in the study of this system to obtain a non-Hermitian invariant $I_H(x,y,t)$ as well as the time-dependent Dyson map $\eta(t)$. Taking (\ref{target}) as the target Hamiltonian satisfying the TDSE
\begin{equation}\label{TDSEtar}
	H(x,y,t)\Phi(x,y,t) = \imath\hbar \partial_t\Phi(x,y,t),
\end{equation}
we choose for our reference Hamiltonian a time-independent version of (\ref{target}) given by
\begin{equation}\label{ref}
	H_0(\chi,\upsilon) = \frac{\alpha}{2}\left(p_\chi^2+\chi^2\right)+\frac{\beta}{2}\left(p_\upsilon^2+\upsilon^2\right)+\imath\Lambda \chi \upsilon, \quad \alpha,\beta,\Lambda \in \mathbb{R},
\end{equation}
satisfying the TDSE
\begin{eqnarray}\label{TDSEref}
	H_0(\chi,\upsilon)\Psi(\chi,\upsilon,\tau) = \imath \hbar \partial_\tau\Psi(\chi,\upsilon,\tau).
\end{eqnarray}
We simplify the general functional dependences (\ref{genpoint}) to
\begin{equation}\label{funcdep}
	\chi = \chi(y,t), \quad \upsilon = \upsilon(x,t), \quad  \tau = \tau(t), \quad \Psi = A(x,y,t) \Phi(x,y,t).
\end{equation}
and convert all the partial derivatives in the TDSE from the  $(\chi,\upsilon,\tau)$ to the $(x,y,t)$-variables obtaining the point transformed differential equation
\begin{equation}\label{pointtrans}
	\imath \hbar \Phi_t +\frac{\hbar^2}{2}\frac{\beta\tau_t}{\upsilon_x^2}\Phi_{xx}+\frac{\hbar^2}{2}\frac{\alpha\tau_t}{\chi_y^2}\Phi_{yy}+B_{0,x}(x,y,t)\Phi_x+B_{0,y}(x,y,t)\Phi_y-V_0(x,y,t)\Phi = 0,
\end{equation}
with
\begin{align}
	B_{0,x}(x,y,t) &= -\frac{\imath \hbar  \upsilon_t}{\upsilon_x}+\frac{\hbar^2}{2}\frac{\tau_t}{\upsilon_x^2}\left(\frac{2 \beta  A_x}{A}-\frac{\beta  \upsilon _{xx}}{\upsilon _x}-\frac{\alpha  \upsilon _x \upsilon _{yy}}{\chi _y^2}\right), \label{pt1}\\
	B_{0,y}(x,y,t) &= - \frac{\imath\hbar \chi_t}{\chi_y}+\frac{\hbar^2}{2}\frac{\tau_t}{\chi_y^2}\left(\frac{2 \alpha  A_y}{A}-\frac{\beta  \chi _y \chi _{xx}}{\upsilon _x^2}-\frac{\alpha  \chi _{yy}}{\chi _y}\right),\\
	V_0(x,y,t) &= \frac{\tau_t}{2}\left(\beta\upsilon^2+2i\Lambda\chi\upsilon+\alpha\chi^2\right)-\imath \hbar \left(\frac{A_t}{A
	}-\frac{A_x\upsilon_t}{A\upsilon_x}-\frac{A_y\chi_t}{A\chi_y}\right) \label{pt3}\\&~-\frac{\hbar^2\tau_t}{2A}\left[\frac{\alpha}{\chi_y^2}\left(A_{y,y}-\frac{A_x \upsilon _{y,y}}{\upsilon _x}-\frac{A_y \chi _{yy}}{\chi _y}\right)+\frac{\beta}{\upsilon_x^2}\left(A_{xx}-\frac{A_x \upsilon _{xx}}{\upsilon _x}-\frac{A_y \chi _{xx}}{\chi _y}\right)\right].
	&\notag 
\end{align}
The motivation behind the first two assumptions in (\ref{funcdep}) is dictated by the form of the target differential equation and the fact that there are no $\Phi_{xy}$ terms present. There are in fact other choices to ensure this term to vanish, but for now this simplification is adequate for the construction of the point transformation. Similarly, the last factorization property in (\ref{funcdep}) ensures that there are no nonlinear $\Phi_x^2$ and $\Phi_y^2$ terms. Having $\tau$ only being a function of $t$ is to reduce the complexity of the calculation. 
\subsubsection{Point transformation $\Gamma$ for the target Hamiltonian $H(x,y,t)$}
The target Hamiltonian is chosen to be the one in (\ref{target}) with associated TDSE
\begin{equation}\label{targetpoint}
	\imath\hbar \Phi_t + \frac{\hbar^2}{2}a(t)\Phi_{xx}+\frac{\hbar^2}{2}b(t)\Phi_{yy}-\left(\frac{1}{2}a(t)x^2+\frac{1}{2}b(t)y^2+\imath\lambda(t) xy\right)\Phi = 0.
\end{equation}
Comparing this equation directly with the differential equation (\ref{pointtrans}) leads to the five constraints
\begin{eqnarray}\label{cons}
	a(t) = \frac{\beta\tau_t}{\upsilon_x^2}, \quad b(t) &=& \frac{\alpha\tau_t}{\chi_y^2}, \quad B_{0,x}  (x,y,t) = 0, \quad B_{0,y}(x,y,t) = 0,\\
	V_0(x,y,t) &=& \frac{1}{2}\left(a(t)x^2  +2\imath\lambda(t)x y+b(t)y^2 \right). \label{cons2}
\end{eqnarray}
To guarantee that $a(t)$ is different from $b(t)$ we introduce the three real valued functions $\mu(t)$, $\sigma(t)$ and $r(t)$, and solve the first two constraints with
\begin{equation}
	\upsilon(x,t) = \mu(t) x + \gamma_1(t), \quad \chi(y,t) = \sigma(t) y +\gamma_2(t), \quad \tau(t) = \int^tr(s) ds,
\end{equation}
where $\gamma_1(t)$ and $\gamma_2(t)$ and real valued constants of integration. Using these expressions in the third and fourth constraints of (\ref{cons}) yields the two equations
\begin{equation}
	\hbar\frac{\alpha r A_x}{A\sigma^2} = \imath \left(\frac{\sigma_t}{\sigma}x+\frac{(\gamma_2)_t}{\sigma}\right), \qquad \text{and} \qquad
	\hbar\frac{\beta r A_y}{A\mu^2} = \imath \left(\frac{\mu_t}{\mu}y+\frac{\gamma_t}{\mu}\right).
\end{equation}
These two equations are solved by
\begin{equation}
	A(x,y,t) = \exp\left\lbrace \frac{\imath}{2\alpha\beta\hbar r}\left[\beta y \mu \left(2\gamma_t+y\mu_t\right)+\alpha x \sigma\left(2(\gamma_2)_t+x\sigma_t\right) \right]+\delta(t)\right\rbrace,
\end{equation}
where $\delta(t)$ is a real valued function resulting from integration. We proceed next with these expressions in the fifth and final constraint in (\ref{cons2}), yielding
\begin{align}
	&\frac{1}{2}\left[\imath\hbar \left(2\delta_t+\frac{\mu_t}{\mu}+\frac{\sigma_t}{\sigma}\right)-r\left(\alpha\gamma_1^2+2\imath\lambda\gamma_1\gamma_2+\beta\gamma_2^2\right)+\frac{1}{r}\left(\frac{(\gamma_1)_t^2}{\alpha}+\frac{(\gamma_2)_t^2}{\beta}\right)\right] \\&-\frac{\sigma}{\beta r^2}\left[\beta r^3\left(\imath\Lambda \gamma_1+\beta\gamma_2\right)-r_t(\gamma_2)_t+r(\gamma_2)_{tt}\right]x+\imath\left(\lambda-\Lambda r\sigma\mu\right)xy\notag\\&-\frac{\mu}{\alpha r^2}\left[\alpha r^3\left(\imath\Lambda \gamma_2+\alpha\gamma_1\right)-r_t(\gamma_1)_t+r(\gamma_1)_{tt}\right]y+\frac{1}{2} \left[\frac{\sigma  \left(r_t \sigma _t-r \sigma _{\text{tt}}\right)}{\beta  r^2}-\frac{\beta  r \left(\sigma ^4-1\right)}{\sigma ^2}\right]x^2\notag\\&+\frac{1}{2} \left[\frac{\mu  \left(r_t \mu _t-r \mu _{\text{tt}}\right)}{\alpha  r^2}-\frac{\alpha  \left(\mu ^4-1\right) r}{\mu ^2}\right]y^2 = 0.\notag
\end{align}
We notice immediately that the $xy$-dependent term vanishes for 
\begin{equation}
	\lambda = \Lambda r \sigma \mu.
\end{equation}
We set $\gamma_1 = \gamma_2 = 0$ ensuring that the $x$ and $y$ dependent terms vanish. The $x$ and $y$ independent terms vanish for
\begin{equation}
	\delta(t) = c_1-\frac{1}{2}\ln(\mu\sigma),
\end{equation}
where $c_1$ is a real time-independent integration constant. Finally we notice that the coefficient functions for the $x^2$ and $y^2$ terms amount to two independent dissipative Ermakaov-Pinney equations \cite{Ermakov,Pinney}
\begin{equation}
	\sigma_{tt}-\frac{r_t}{r}\sigma_t+\beta^2r^2\sigma^2 = \frac{\beta^2r^2}{\sigma^3}, \qquad
	\text{and}
	\qquad
	\mu_{tt}-\frac{r_t}{r}\mu_t+\alpha^2r^2\mu^2=\frac{\alpha^2r^2}{\mu^3}.
\end{equation}
The solutions to these two equations are
\begin{eqnarray}
	\sigma(t) &=& \sqrt{\sqrt{1+c_2^2}+c_2\cos\left[2\beta\int^t r(s) ds\right]},\\
	\mu(t) &=& \sqrt{\sqrt{1+c_3^2}+c_3\cos\left[2\alpha\int^t r(s) ds\right]},
\end{eqnarray}
respectively, where $c_2$ and $c_3$ are constants.

With this we have completed the construction of the point transformation $\Gamma$ and we can use it next to obtain a non-Hermitian invariant for the target Hamiltonian (\ref{target}) as well as the time-dependent Dyson map. These calculations will be carried in terms of the generators of the generators of the $sp(4)$ Lie algebra and so it is instructive here to present how the point transformation acts on the individual generators. 

\subsubsection{LR-invariants from $\Gamma$}
We now use the point transformation $\Gamma$ defined by the equations (\ref{p1})-(\ref{p10}) and (\ref{p11}) to obtain the non-Hermitian invariant for the target Hamiltonian (\ref{target}). Acting directly on the reference Hamiltonian (\ref{ref}) and not the TDSE, we obtain
\begin{align}\label{nonI}
	I_H(t) &= \frac{\beta}{2\sigma^2}(J_3+J_1+J_0+Q_2)+\frac{\alpha}{2\mu^2}(J_0+Q_2-J_3-K_1)+\frac{\sigma_t}{r\sigma}(K_2-Q_1) \\ \notag &+\frac{\mu_t}{r \mu}(K_2+Q_1)+\frac{1}{2}\left(\frac{\sigma_t^2}{\beta r^2}+\beta\sigma^2\right)(J_3-K_1+J_0-Q_2)\\\notag&+\frac{1}{2}\left(\frac{\mu_t^2}{\alpha r^2}+\alpha\mu^2\right)(K_1-J_3+J_0-Q_2)+i\Lambda\sigma\mu(J_1+K_3).
\end{align}
We have verified that this expression for the invariant $I_H(t)$ does indeed satisfy the Lewis-Riesenfeld equation (\ref{LRdef}).
\subsubsection{Time-dependent Dyson map from $\Gamma$}
Next we use the point transformation $\Gamma$ to obtain the time-dependent Dyson map for the target Hamiltonian (\ref{target}). To do so we must first determine the time-independent Dyson map for the reference Hamiltonian (\ref{ref}). We make the following Ansatz for the time-independent Dyson map
\begin{equation}
	\eta = \exp[\kappa_1(Q_3-J_2)+\kappa_2(Q_3+J_2)], \quad \kappa_1,\kappa_2\in\mathbb{R},
\end{equation}
and act adjointly with it on the reference Hamiltonian (\ref{ref}). The requirement for the non-Hermitian terms to vanish to ensure that we are left with a  Hermitian Hamiltonian results in the two constraining equations 
\begin{align}\label{eq1}
	2\Lambda\cos[2 \sqrt{\kappa_1\kappa_2}] &=\frac{(\alpha+\beta)(\kappa_1+\kappa_2)\sin[2\sqrt{\kappa_1\kappa_2}]}{\sqrt{\kappa_1\kappa_2}},
\end{align}
and
\begin{equation}
	2\Lambda\cos[2 \sqrt{\kappa_1\kappa_2}] =\frac{(\alpha-\beta)(\kappa_1-\kappa_2)\sin[2\sqrt{\kappa_1\kappa_2}]}{\sqrt{\kappa_1\kappa_2}}.
\end{equation}
These equations are easily solved for $\kappa_1$ and $\kappa_2$ 
\begin{equation}
	\kappa_1 = \frac{1}{2}\sqrt{\frac{\alpha}{\beta}}\arctanh\left[\frac{2\sqrt{\alpha\beta}\Lambda}{\alpha^2-\beta^2}\right] \qquad \text{and} \qquad \kappa_2 =- \frac{1}{2}\sqrt{\frac{\beta}{\alpha}}\arctanh\left[\frac{2\sqrt{\alpha\beta}\Lambda}{\alpha^2-\beta^2}\right] .
\end{equation}
The resulting Hermitian Hamiltonian is then given by
\begin{eqnarray}\label{h0}
	h_0&=&\eta H_0\eta^{-1} \\ &=& \frac{(\alpha+\beta)\Delta-(\alpha-\beta)^3}{4\alpha\beta}J_3+\frac{(\beta-\alpha)\Delta+(\alpha+\beta)^3}{4\alpha\beta}J_0  \notag \\ && +\frac{(\alpha^2-\beta^2-\Delta)}{4\alpha\beta}\left[(\alpha+\beta)K_1-(\alpha-\beta)Q_2\right], \notag
\end{eqnarray}
where we introduced the abbreviation
\begin{equation}
	\Delta := \sqrt{(\alpha^2-\beta^2)^2-4\alpha\beta\Lambda^2}
\end{equation}

Now that we have determined the time-independent Dyson map, we can act on it with the point transformation $\Gamma$ to obtain
\begin{equation}\label{etat}
\Gamma	(\eta) \rightarrow \eta(t) = \exp[\kappa_2\frac{\mu}{\sigma}(Q_3-J_2)+\kappa_1\frac{\sigma}{\mu}(Q_3+J_2)+\frac{\beta \kappa_1\sigma\mu_t+\alpha\kappa_2\mu\sigma_t}{\alpha\beta r}(K_3+J_1)].
\end{equation}
Acting adjointly with this map on the time-dependent non-Hermitian invariant (\ref{nonI}) does indeed produce a Hermitian invariant given by
\begin{align}\label{herminv}
	I_h(t) =& \eta(t)I_H(t)\eta(t)^{-1} = \frac{1}{4}\left\lbrace\frac{2\alpha}{\mu^2}(J_0-J_3-K_1+Q_2)+\frac{2\beta}{\sigma^2}(J_0+J_3+K_1+Q_2)\right. \\
	\notag &+\frac{4\mu_t}{r\mu}(K_2+Q_1) +\frac{1}{\alpha}\left[(\alpha^2+\beta^2+\Delta)\mu^2+\frac{2\mu_t^2}{r^2}\right](J_0-J_3+K_1-Q_2)\\\notag &\left.+\frac{4\sigma_t}{r\sigma}(K_2-Q_1)+\frac{1}{\beta}\left[(\alpha^2+\beta^2-\Delta)\sigma^2+\frac{2\sigma_t^2}{r^2}\right](J_0+J_3-K_1-Q_2)\right\rbrace.
\end{align}
Given that this invariant is Hermitian we conclude that equation (\ref{etat}) is a Dyson map for the target Hamiltonian (\ref{target})
\subsubsection{Time-dependent Hermitian Hamiltonian from $\Gamma$}
We now wish to determine the time-dependent Hermitian Hamiltonian which is related to the target Hamiltonian (\ref{target}) via the time-dependent Dyson equation
\begin{equation}
	h(t)=\eta (t)H(t)\eta ^{-1}(t)+i\hbar \partial _{t}\eta (t)\eta (t)^{-1}. \label{TDDE}
\end{equation}%
 In general this equation is difficult to solve due the presence of the time-derivative terms on the Dyson map. Here we show that alternatively we can instead use the point transformation to find $\eta(t)$ and $h(t)$ directly. 

We start by acting with the point transformation $\Gamma$ on the time-independent Hermitian Hamiltonian constructed in (\ref{h0}) which leads to a time-dependent Hermitian invariant for this Hamiltonian, $\Gamma(h_0) \rightarrow I_h(t)$. We verify that the result is indeed the Hermitian invariant given in equation (\ref{herminv}). Subsequently, corresponding to the last step in our scheme outlined in section 3.2.1, we act with $\Gamma$ on the entire TDSE satisfied by $h_0$ and identify from the resulting equation the time-dependent Hermitian Hamiltonian 
\begin{align}
	h(t) &= \frac{r \alpha}{\mu^2}(J_0-J_3)+\frac{r\beta}{\sigma^2}(J_0+J_3)-\frac{r\mu^2}{4\alpha}(\alpha^2-\beta^2-\Delta)(J_0-J_3+K_1-Q_2) \\
	\notag &-\frac{r\sigma^2}{4\beta}(\beta^2-\alpha^2+\Delta)(J_0+J_3-K_1-Q_2).
\end{align} 
This Hamiltonian is indeed the one on the right hand side in (\ref{TDDE}) and is indeed the Hermitian Hamiltonian for the invariant constructed in (\ref{herminv}).
\section{Conclusions}
As our two main results we showed first of all that the previously observed feature that LR-invariants can be obtained directly from the action of point transformation on a time dependent Hamiltonian generalised from one to two dimensions. Furthermore we showed that time-dependent Dyson maps and time-dependent Hermitian counterparts can also be obtained by means of the action of the point transformation on the corresponding time-independent Dyson maps and TDSE, respectively. Hence this approach completely bypasses to solve the time-dependent Dyson equation (\ref{TDDE}) directly.  

Here we are content with a proof of concept, but naturally it would be interesting to explore this scheme further, as for instance constructing the metric operator from the Dyson map and investigate its properties, start with different types of reference Hamiltonians, construct all related eigenfunctions and explore the energy spectra in all ${\cal PT}$-regimes. 

Regarding the model itself our analysis has clearly demonstrated that it is extremely advantageous to express the model entirely in terms of the generators of the $sp(4)$-Lie algebra, which not only greatly simplifies the computations when compared to this in coordinate or momentum space, but also provides a more systematic framework.

\medskip

\noindent \textbf{Acknowledgments:} RT is supported by EPSRC grant EP/W522351/1.

\begin{appendices}
\section{Two dimensional point transformations} 
In this appendix we provide the details on how to derive central equations (\ref{pointtrans}) and (\ref{pt1}) to (\ref{pt3}) given in the main body of the paper. 

We wish to determine the point transformation which maps the TDSE (\ref{TDSEref}) for the reference Hamiltonian to that of the target Hamiltonian (\ref{TDSEtar}). The wavefunctions $\Psi$ and $\Phi$ are implicit functions of the variables $(\chi,\upsilon,\tau)$ and $(x,y,t)$ respectively with assumed functional dependence 
\begin{equation}
	\chi = \chi(x,y,t), \quad \upsilon = \upsilon(x,y,t), \quad 	\tau = \tau(x,y,t), \quad \Psi = R(x,y,t,\Phi(x,y,t)).
\end{equation}
We first compute the total derivatives of $\Psi$ with respect to the variables $x$, $y$ and $t$
\begin{align}
	\frac{d\Psi}{dx} &= \Psi_\chi\chi_x+\Psi_\upsilon\upsilon_x+\Psi_\tau\tau_x = R_\Phi\Phi_x +R_x\\
	\frac{d\Psi}{dy} &= \Psi_\chi\chi_y+\Psi_\upsilon\upsilon_y+\Psi_\tau\tau_y = R_\Phi\Phi_y+R_y \\
	\frac{d\Psi}{dt} &= \Psi_\chi\chi_t+\Psi_\upsilon\upsilon_t+\Psi_\tau\tau_t = R_\Phi\Phi_t+R_t.
\end{align}
We solve this system of equations for the unknown functions $\Psi_\chi$, $\Psi_\upsilon$ and $\Psi_\tau$ obtaining
\begin{align}\label{A5}
	\Psi_\chi &= \frac{1}{J}\left\lbrace R_{\Phi } \left[\upsilon _y \left(\Phi _t \tau _x-\tau _t \Phi _x\right)+\upsilon _x \left(\tau _t \Phi _y-\Phi _t \tau _y\right)+\upsilon _t \left(\Phi _x \tau _y-\tau _x \Phi _y\right)\right]\right.\\ \notag&\quad\quad~~\left.R_y \left(\tau _t \upsilon _x-\upsilon _t \tau _x\right)+R_t \left(\tau _x \upsilon _y-\upsilon _x \tau _y\right)+R_x \left(\upsilon _t \tau _y-\tau _t \upsilon _y\right)\right\rbrace, \\
	\Psi_\upsilon &= \frac{1}{J}\left\lbrace R_{\Phi } \left[\chi _y \left(\tau _t \Phi _x-\Phi _t \tau _x\right)+\chi _t \left(\tau _x \Phi _y-\Phi _x \tau _y\right)+\chi _x \left(\Phi _t \tau _y-\tau _t \Phi _y\right)\right] \right.  \\\notag&\quad\quad~~\left.+R_y \left(\chi _t \tau _x-\tau _t \chi _x\right)+R_x \left(\tau _t \chi _y-\chi _t \tau _y\right)+R_t \left(\chi _x \tau _y-\tau _x \chi _y\right)\right\rbrace, \\\label{A6}
	\Psi_\tau &= \frac{1}{J}\left\lbrace R_{\Phi } \left[\chi _y \left(\Phi _t \upsilon _x-\upsilon _t \Phi _x\right)+\chi _x \left(\upsilon _t \Phi _y-\Phi _t \upsilon _y\right)+\chi _t \left(\Phi _x \upsilon _y-\upsilon _x \Phi _y\right)\right]\right.\\\notag&\quad\quad~~\left.+R_y \left(\upsilon _t \chi _x-\chi _t \upsilon _x\right)+R_t \left(\upsilon _x \chi _y-\chi _x \upsilon _y\right)+R_x \left(\chi _t \upsilon _y-\upsilon _t \chi _y\right)\right\rbrace, 
\end{align}
where 
\begin{equation}
	J = \chi _t \left(\tau _x \upsilon _y-\upsilon _x \tau _y\right)+\chi _x \left(\upsilon _t \tau _y-\tau _t \upsilon _y\right)+\chi _y \left(\tau _t \upsilon _x-\upsilon _t \tau _x\right)
\end{equation}
is the Jacobian. 
We also compute the second derivatives of $\Psi$ with respect to $x$ and $y$ obtaining
\begin{eqnarray}\label{A9}
	\frac{d^2\Psi}{dx^2} &=& 2 \tau _x \left(\upsilon _x \Psi _{\upsilon ,\tau }+\chi _x \Psi _{\chi ,\tau }\right)+\tau _x^2 \Psi _{\tau ,\tau }+\Psi _{\tau } \tau _{x,x}+\Psi _{\upsilon } \upsilon _{x,x}+\upsilon _x^2 \Psi _{\upsilon ,\upsilon } \\\notag &&~+2 \upsilon _x \chi _x \Psi _{\chi ,\upsilon }+\chi _x^2 \Psi _{\chi ,\chi }+\Psi _{\chi } \chi _{x,x}, \\\frac{d^2\Psi}{dx^2} & =& \Phi _x^2 R_{\Phi ,\Phi }+2 \Phi _x R_{\Phi ,x}+R_{\Phi } \Phi _{x,x}+R_{x,x}, \label{A10}\\
	\frac{d^2\Psi}{dy^2} &=& 2 \tau _y \left(\upsilon _y \Psi _{\upsilon ,\tau }+\chi _y \Psi _{\chi ,\tau }\right)+\tau _y^2 \Psi _{\tau ,\tau }+\Psi _{\tau } \tau _{y,y}+\Psi _{\upsilon } \upsilon _{y,y}+\upsilon _y^2 \Psi _{\upsilon ,\upsilon }\\\notag &&~+2 \upsilon _y \chi _y \Psi _{\chi ,\upsilon }+\chi _y^2 \Psi _{\chi ,\chi }+\Psi _{\chi } \chi _{y,y}, \\
	\frac{d^2\Psi}{dy^2}&=&\Phi _y^2 R_{\Phi ,\Phi }+2 \Phi _y R_{\Phi ,y}+R_{\Phi } \Phi _{y,y}+R_{y,y}.\label{A12}
\end{eqnarray}
To remove the nonlinear terms present in (\ref{A10}) and (\ref{A12}) we factorise the wavefunction as
\begin{equation}
	\Psi = A(x,y,t)\Phi,
\end{equation}
such that $R_{\Phi,\Phi} = 0$. To greatly reduce the difficulty of the calculation we also choose to simplify the functional dependence of $\tau$ to
\begin{equation}
	\tau = \tau(t).
\end{equation}
We make one final simplification on the functional dependence of $\chi$ and $\upsilon$ to ensure there are no $\Phi_{XY}$ terms present in the resulting point transformed differential equation, that being
\begin{equation}
	\chi = \chi(y,t) \qquad \text{and} \qquad \upsilon = \upsilon(x,t).    
\end{equation}
We stress here though that this a choice, and there are other simplifications which would lead to the same outcome. 

With all of the simplifications we may now use equations (\ref{A5})-(\ref{A6}) and (\ref{A9})-(\ref{A12}) to obtain 
\begin{eqnarray}
	\Psi_\chi &=& \frac{1}{\chi_y}\left(\Phi  A_y+A \Phi _y\right), \\
	\Psi_\upsilon &=& \frac{1}{\upsilon_x}\left(\Phi  A_x+A \Phi _x\right), \\
	\Psi_{\chi,\chi} &=& \frac{1}{\chi_y^3}\left[A_y \left(2 \chi _y \Phi _y-\Phi  \chi _{y,y}\right)+\chi _y \left(\Phi  A_{y,y}+A \Phi _{y,y}\right)-A \Phi _y \chi _{y,y}\right],\\
	\Psi_{\upsilon,\upsilon} &=& \frac{1}{\upsilon_x^3}\left[A_x \left(2 \upsilon _x \Phi _x-\Phi  \upsilon _{x,x}\right)+\upsilon _x \left(\Phi  A_{x,x}+A \Phi _{x,x}\right)-A \Phi _x \upsilon _{x,x}\right],\\
	\Psi_\tau &=& \frac{1}{\tau_t\chi_y\upsilon_x}\left[\chi _y \left(\upsilon _x \left(\Phi  A_t+A \Phi _t\right) -\Phi  A_x \upsilon _t-A \upsilon _t \Phi _x\right)-\Phi  A_y \chi _t \upsilon _x-A \chi _t \upsilon _x \Phi _y\right]. \,\,\,\,\,\,\quad \,\,
\end{eqnarray}
Substituting these expressions into the TDSE for the reference Hamiltonian (\ref{TDSEref}) yields the point transformed differential equation given in (\ref{pointtrans}).

\section{The point transformation $\Gamma$ acting on the $sp(4)$ generators}
In the appendix we present the details on how the point transformation acts on the generators of the $sp(4)$- Lie algebra. Denoting the generators for the reference Hamiltonian with primes as
\begin{eqnarray}
	J_0' &=& \frac{1}{4} \left( p_\chi^2 +p_\upsilon^2 +  \chi^2 +  \upsilon^2   \right), \quad \label{rep21}\\
	J_1' &=& \frac{1}{2} \left( \chi \upsilon+ p_\chi p_\upsilon    \right), \quad
	J_2' = \frac{1}{2} \left(\chi p_\upsilon - \upsilon p_\chi   \right), \quad 
	J_3' = \frac{1}{4} \left( p_\chi^2 - p_\upsilon^2 + \chi^2  - \upsilon^2   \right), \quad\\
	Q_1' &=& \frac{1}{2} \left( \upsilon p_\upsilon -  \chi p_\chi   \right), \quad 
	Q_2' = \frac{1}{4} \left( p_\chi^2  + p_\upsilon^2 - \chi^2 - \upsilon^2   \right), \quad
	Q_3' = \frac{1}{2} \left( \chi p_\upsilon +  \upsilon p_\chi   \right), \quad \\
	K_1' &=& \frac{1}{4} \left( p_\chi^2  - p_\upsilon^2 - \chi^2 - \upsilon^2   \right), \quad
	K_2' = \frac{1}{2} \left( \chi p_\chi +   p_\upsilon\upsilon   \right), \quad
	K_3' = \frac{1}{2} \left( \chi \upsilon -  p_\chi p_\upsilon   \right), \label{rep22}
\end{eqnarray}
the action of the point transformation $\Gamma$ on them are then computed to
\begin{eqnarray}\label{p1}
	&& J_0' \xrightarrow{\Gamma} \frac{1}{4}\left[\frac{J_0-J_3-K_1+Q_2}{\mu ^2}+\frac{J_0+J_3+K_1+Q_2}{\sigma ^2}+\frac{2 \left(K_2-Q_1\right) \sigma _t}{\beta  r \sigma }\right.\\ \notag 
	&& \qquad \left.+\frac{\left(J_0+J_3-K_1-Q_2\right) \left(\beta ^2 r^2 \sigma ^2+\sigma _t^2\right)}{\beta ^2 r^2}+\frac{2 \left(K_2+Q_1\right) \mu _t}{\alpha  \mu  r}.\right. \\ \notag 
	&& \qquad \left.+\frac{\left(J_0-J_3+K_1-Q_2\right) \left(\alpha ^2 \mu ^2 r^2+\mu _t^2\right)}{\alpha ^2 r^2}\right] \\
	&& J_1' \xrightarrow{\Gamma} \frac{J_1-K_3}{2 \mu  \sigma }+\frac{1}{2} \left(J_1+K_3\right) \left(\mu  \sigma +\frac{\mu _t \sigma _t}{\alpha  \beta  r^2}\right)+\frac{\left(Q_3-J_2\right) \mu _t}{2 \alpha  r \sigma }+\frac{\left(J_2+Q_3\right) \sigma _t}{2 \beta  \mu  r}, \\
	&& J_2' \xrightarrow{\Gamma} \frac{\left(J_1+K_3\right) \left(\alpha  \mu  \sigma _t-\beta  \sigma  \mu _t\right)}{2 \alpha  \beta  r}+\frac{\mu  \left(Q_3-J_2\right)}{2 \sigma }-\frac{\sigma  \left(J_2+Q_3\right)}{2 \mu }, \\
	&& J_3' \xrightarrow{\Gamma} \frac{1}{2}\left[\frac{J_0-J_3-K_1+Q_2}{\mu ^2}+\frac{2 \left(K_2+Q_1\right) \mu _t}{\alpha  \mu  r}-\frac{J_0+J_3+K_1+Q_2}{\sigma ^2} \right. \\\notag && \qquad +\frac{\left(J_0-J_3+K_1-Q_2\right) \left(\alpha ^2 \mu ^2 r^2+\mu _t^2\right)}{\alpha ^2 r^2}+\frac{2 \left(Q_1-K_2\right) \sigma _t}{\beta  r \sigma }\\\notag && \qquad \left.-\frac{\left(J_0+J_3-K_1-Q_2\right) \left(\beta ^2 r^2 \sigma ^2+\sigma _t^2\right)}{\beta ^2 r^2}\right],\\
	&& Q_1' \xrightarrow{\Gamma}\frac{\sigma  \sigma _t \left(J_0+J_3-K_1-Q_2\right)}{2 \beta  r} -\frac{\mu  \mu _t \left(J_0-J_3+K_1-Q_2\right)}{2 \alpha  r}-Q_1, \\
	&& Q_2' \xrightarrow{\Gamma} \frac{1}{4}\left[\frac{J_0-J_3-K_1+Q_2}{\mu ^2}+\frac{J_0+J_3+K_1+Q_2}{\sigma ^2}+\frac{2 \left(K_2+Q_1\right) \mu _t}{\alpha  \mu  r}\right. \\ \notag && \qquad +\frac{2 \left(K_2-Q_1\right) \sigma _t}{\beta  r \sigma }-\frac{\left(J_0-J_3+K_1-Q_2\right) \left(\alpha ^2 \mu ^2 r^2-\mu _t^2\right)}{\alpha ^2 r^2}\\ \notag && \qquad \left.-\frac{\left(J_0+J_3-K_1-Q_2\right) \left(\beta ^2 r^2 \sigma ^2-\sigma _t^2\right)}{\beta ^2 r^2}\right], \\
	&& Q_3' \xrightarrow{\Gamma} \frac{\left(J_1+K_3\right) \left(\alpha  \mu  \sigma _t+\beta  \sigma  \mu _t\right)}{2 \alpha  \beta  r}+\frac{\mu  \left(Q_3-J_2\right)}{2 \sigma }+\frac{\sigma  \left(J_2+Q_3\right)}{2 \mu }, \\
  &&	K_1' \xrightarrow{\Gamma} \frac{1}{4}\left[\frac{J_0-J_3-K_1+Q_2}{\mu ^2}-\frac{J_0+J_3+K_1+Q_2}{\sigma ^2}+\frac{2 \left(K_2+Q_1\right) \mu _t}{\alpha  \mu  r}\right. \\ \notag && \qquad +\frac{2 \left(Q_1-K_2\right) \sigma _t}{\beta  r \sigma }-\frac{\left(J_0-J_3+K_1-Q_2\right) \left(\alpha ^2 \mu ^2 r^2-\mu _t^2\right)}{\alpha ^2 r^2} \\ \notag && \qquad\left.+\frac{\left(J_0+J_3-K_1-Q_2\right) \left(\beta ^2 r^2 \sigma ^2-\sigma _t^2\right)}{\beta ^2 r^2}+\frac{2 \left(Q_1-K_2\right) \sigma _t}{\beta  r \sigma }\right], \\
&&	K_2' \xrightarrow{\Gamma} \frac{\mu  \mu _t \left(J_0-J_3+K_1-Q_2\right)}{2 \alpha  r}+\frac{\sigma  \sigma _t \left(J_0+J_3-K_1-Q_2\right)}{2 \beta  r}+K_2, \\\label{p10}
&&	K_3' \xrightarrow{\Gamma} -\frac{J_1-K_3}{2 \mu  \sigma }+\frac{1}{2} \left(J_1+K_3\right) \left(\mu  \sigma -\frac{\mu _t \sigma _t}{\alpha  \beta  r^2}\right)-\frac{\left(Q_3-J_2\right) \mu _t}{2 \alpha  r \sigma }-\frac{\left(J_2+Q_3\right) \sigma _t}{2 \beta  \mu  r}. \qquad \,\,\,\,\,
\end{eqnarray}
The time-derivative $\Psi_\tau$ when expressed in terms of the Lie algebraic generators then transforms as
\begin{align}\label{p11}
	i \hbar \tau_t \Psi_\tau &\xrightarrow{\Gamma} i \hbar \Psi_t +r \left(J_0-J_3+K_1-Q_2\right) \left(\frac{\alpha  \left(\mu ^4-1\right)}{2 \mu ^2}+\frac{\mu _t^2}{2 \alpha  r^2}\right)+\frac{\left(K_2+Q_1\right) \mu _t}{\mu }\\ &+\frac{1}{2} r \left(J_0+J_3-K_1-Q_2\right) \left(\frac{\beta  \left(\sigma ^4-1\right)}{\sigma ^2}+\frac{\sigma _t^2}{\beta  r^2}\right)+\frac{\left(K_2-Q_1\right) \sigma _t}{\sigma }. \notag
\end{align}
\end{appendices}

\newif\ifabfull\abfulltrue


\end{document}